\documentclass{llncs}
\usepackage{graphicx}
\usepackage[hyphens]{url}
\usepackage{breqn}
\usepackage{hyperref} 
\usepackage{hyphenat}
\usepackage{subfig}
\begin{document}
\pagestyle{headings}  % switches on printing of running heads
\title{Humans of Simulated New York – HOSNY: \\
	an exploratory comprehensive model of city life}
\titlerunning{Humans of Simulated New York}  
\author{Francis Tseng \inst{1} \and Fei Liu \inst{1} \and Bernardo Alves Furtado \inst{2, 3}}
\authorrunning{Tseng, Liu \and Furtado} % abbreviated author list (for running head)
\tocauthor{Francis Tseng, Fei Liu and Bernardo Alves Furtado}
\institute{Public Science Agency and NEW INC\\
\email{f@frnsys.com} \\
\email{fei.fei2liu@gmail.com}
\and
	Institute of Applied Economic Research, IPEA\\
\email{bernardo.furtado@ipea.gov.br}\\ 
\and
National Council of Research, CNPq\\
}

\maketitle             

\begin{abstract}
The model presented in this paper experiments with a comprehensive simulant agent in order to provide an exploratory platform in which simulation modelers may try alternative scenarios and participation in policy decision-making. The framework is built in a computationally distributed online format in which users can join in and visually explore the results. Modeled activity involves daily routine errands, such as shopping, visiting the doctor or engaging in the labor market. Further, agents make everyday decisions based on individual behavioral attributes and minimal requirements, according to social and contagion networks. Fully developed firms and governments are also included in the model allowing for taxes collection, production decisions, bankruptcy and change in ownership. The contributions to the literature are manifold. They include (a) a comprehensive model with detailing of the agents and firms' activities and processes and original use of simultaneously (b) reinforcement learning for firm pricing and demand allocation; (c) social contagion for disease spreading and social network for hiring opportunities; and (d) Bayesian networks for demographic-like generation of agents. All of that within a (e) visually rich environment and multiple use of databases. Hence, the model provides a comprehensive framework from where interactions among citizens, firms and governments can be easily explored allowing for learning and visualization of policies and scenarios.
 
\end{abstract}

\section{Introduction}

City life encapsulates the bonuses of agglomeration \cite{fujita,oecd2}. Cities are productive, cheap to build and maintain \cite{bettencourt_west} and even sustainable \cite{glaeser}. However, cities can be extremely complex \cite{bettencourt} and difficult to manage \cite{oecd1}. Essentially, city life is dynamic and evolves in an open system through interactions among its actors: citizens, businesses, governments, and the environment. 

Citizens and tax-paying voters (as well as city employees) may have difficulties to make sense of city life, its dynamics and the extension and nature of the effects of policy decision-making. This understanding is especially relevant when simultaneously considering economic features (the labor and housing markets, production), along with health issues (disease contagion and spreading), individual attributes and choice making (consumption decisions or labor referral, for example). 

Scientists \cite{benevolo,sitte} and artists \cite{baudelaire,dickens} alike have been trying to comprehend and model city life, either trying to understand their mental models \cite{tversky} or human perception of space \cite{cullen}. From the economic viewpoint, early models \cite{medina1,medina2} have tried mixing political thinking, government decision making and computer models in an attempt to anticipate city life developing. 

Over 40 years after such early attempts \cite{medina2}, a number of available resources have become available. These advances include not only modeling techniques, but also scientific computing, economic foundations and visualizing tools. Given such a context, being able to anticipate general outcomes of policymaking while exploring alternative scenarios in a visually rich environment is one of the goals of the proposed model. 

In fact, this paper proposes and discusses a comprehensive agent-based model based on official data for New York City that incorporates citizens and their individual attributes; businesses, their creation, production and interaction; government and the interaction among all elements using social and contagion networks. It explicitly adds to the literature when considering detailed representation of economic, social and visual aspects of the model, to the best of our knowledge not present in the previous literature. The model implement such features using a distributed agent-based implementation that makes use of reinforcement learning and Bayesian networks. 

Besides this introduction, the literature of agent-based modeling, its economic foundation and previous models of city life are discussed in section \ref{lit}. Section \ref{model} details the model, following Grimm et al. \cite{grimm1,grimm2}, presenting an overview of its processes, the available scenarios as well as the data input and setup procedures. Section \ref{outputs} is dedicated to presenting the visual aspects of the model and the interaction with the users and a quick illustrative exploration of the results. The closing section (\ref{final}) discusses the contributions of the model and sets future work. 

\section{Literature}\label{lit}

One of the first applications of agent-based modeling was to investigate segregation within the city \cite{schelling}. In a broader sense, simulation has been used to study system dynamics derived from simple rules \cite{gardner}, such as Conway’s Game of Life. Since then, an abundant amount of theory has been developed \cite{epstein_axtell,epstein} to frame agent-based modeling as a powerful tool in aiding social phenomena understanding. 

Formally, agent-based modeling can be described as a dynamic system that is temporally discrete \cite{epstein_axtell}. An algorithm processes rules of interaction among agents and the environment and vice-versa from one time-step into the next one, so that:

\begin{equation}
Agents^{t+1}=f(Agents^t, Environment^t)
\end{equation}
\begin{equation}
Environment^{t+1}=g(Agents^t, Environment^t)
\end{equation}

Such a simple schemata allows for a formal description of the model in which all the rules $f()$ and $g()$ governing the changes from $t$ to $t+1$ are described, understood, and validated by theory or data. Hence, there is no 'black box'. 

The results of the model, in turn, cannot be anticipated for large, complex simulations. That is exactly why simulations are needed and emergence of properties can be observed \cite{furtado_sako}.

In fact, Gilbert and Terna \cite{gilbert}, following Ostrom \cite{ostrom}, have suggested that computational modeling is a 'third-way' of doing science. Whereas verbal argumentation and mathematics constitute the two original ones, computer simulation or computational modeling represented an alternative.

\begin{quotation}
	Computer programs can be used to model either quantitative theories or qualitative ones. They are particularly good at modelling processes and although non-linear relationships can generate some methodological problems, there is no difficulty in representing them within a computer program \cite[p. 3]{gilbert}. 
\end{quotation}

Joshua M. Epstein (2011) goes a bit further, claims that simulation is complementary, and differs from both deduction and induction. He considers that inductive reasoning departs from empirical observation in order to derive theory and deductive reasoning is based on a theory, which needs to be confirmed in empirical observation. Simulation – in turn – generates (artificial) data through a design process derived from theory and then goes on to analyze such data in an inductively way. In doing so, simulation can uncover mechanisms and understand interactions and dynamics that would otherwise be very difficult to observe or theorize about. 

Simulation has been used in economics both on a simple scope \cite{furtado,hulia,lengnick} or on comprehensive ones \cite{dawid} with both advantages and disadvantages. Simpler models enable the understanding of the general aspects and mechanisms of the problem at hand, the interactions and the considerations of spatial and time dynamics. As simple models, their predictive capacity is low. Complex models focus on replicating nearly the universe of agents aiming at forecasting and demanding detailed modules and large processing capacity. They are more precise when forecasting short to average-term results, but large models are inherently more difficult to understand, and especially more difficult to separate cause and effect, to establish which parameter or mechanism is responsible for which result. 

In practice, economic modeling using agent-based modeling has been applied successfully in energy markets \cite{koesr,li}, economics \cite{tesfatsion}, and financial analysis \cite{battiston,feng,lebaron}. Agent-based modeling and networks have also been used in ecology \cite{glaeser2}, demography \cite{billari} and health analysis \cite{mesquita}. However, not even the large scale economics model, such as EURACE@Unibi \cite{dawid} implement simultaneously so many aspects of city life as the proposed model does. 

Considering city life simulation, the examples in the literature are scarce. We do have a lot of analysis on land-use change and land-use cover 
\cite{filatova,parker,white}, transport and real estate market \cite{waddell} and also work on visualization \cite{battya} and cities and complexity \cite{battyb}. There has also been references to how people spend their time within the city \cite{ettema}, considering the underlying economics. Civilizations were modeled in Dean et al. \cite{dean}. 

Finally, a number of studies focuses on social aspects embedded in urban tissue \cite{axelrod,epstein,wilensky}. City life, however, has been described more in the sense of symbolic values \cite{lynch} and perception, rather than a simulation object.

Thus, our model uses agent-based modeling of city life using a number of different agents and processes in a comprehensive manner, but observing simple models for each process. The model is based on data, but does not replicate actual living personas; rather it creates simulant who would likely exist, giving the observed data. 

\section{The model}\label{model}

\subsection{Purpose and design concepts}

The model was thought as a framework of experimentation of city life based on real data. As such, daily lives activities were included, along with main businesses and their markets. All of the agents interacting via social networks. The main goal of the experimentation is the voting system and the scenario alternatives. 

The details of the model are described below. It includes the setup and initialization procedures, followed by the methodological contributions of the model. Then, the detailing of agents, businesses, government and markets follow. Finally, the alternative scenarios and the necessary data are described. 

Modeling-wise, the model takes advantage of a previously developed and available platform, known as CESS. Cess is short for 'cesspool' in the sense of a temporary container and storage holder. CESS includes support for distributed simulations. \footnote{CESS is available at https://github.com/frnsys/cess} There are three components to the distributed simulation support:

\begin{itemize}
	\item the Workers, which perform simulation steps locally
	\item the Arbiter, which manages workers and mediates their communication
	\item the Cluster, which communicates to the Arbiter (i.e. sends commands to the cluster) 
\end{itemize}

\subsection{Setup and initialization}

The model is simulated using a distributed web based app using \texttt{Celery}, a task assigner library, along with \texttt{REDIS}, a server messenger that handles the queuing process. Thus, the interface with the modeler is made through an internet browser @localhost. \footnote{The model is available for download at GitHub: https://github.com/frnsys/hosny}

Following such a procedure means that \texttt{app} tasks and routes organize the calling of the processes in the model. In short, the simulation starts with a city setup and run monthly showing online results in the user interface.

\subsection{Methodological contributions}

The model here presented uses $Q$ Learning \cite{chan} as a reinforcement technique in which agents account for past results awarding positive values for correct decisions and negative, punitive deductions for wrong ones. Therefore, agents are continuously updating current decision-making processes. Specifically, firms in the model use $Q$ Learning technique to decide on production levels and establish prices. 

A Bayesian network is applied to official data in order to generate the agents of the model. In doing so, the model can be populated with agents that reflect observed correlations. For example, given a Chinese, middle-aged New Yorker, what neighborhood are they likely to live in, and what is their estimated income? Simulant agents obtained in such a way are not actually present in the data, but it does mirror patterns of the original that in the sense that it is a plausible person. 

A Bayesian network ("Bayes net" for short) is a directed acyclic graph representing conditional dependencies between random variables. Each node in the graph is a random variable, and edges represent conditional dependencies; that is a random variable is conditioned on its parent node(s). Bayes nets provide a compact way of representing such distributions without needing to specify the full joint distribution.

There are various methods for sampling from a Bayes net, one of which is prior sampling. Prior sampling involves sampling a node $N$ by sampling its root node, then sampling that node’s children, conditioned on the root node sample, and repeating until reaching the node $N$.

We were interested in exploring the effect of personal relationships on job attainment, so we used the logistic regression model described in \cite{smith} to generate a social network among the simulated population based on their demographics. The logistic regression model returns a simple probability that two individuals are friends based on race, sex, age, and educational attainment.

\subsection{Process overview and scheduling}

\begin{figure}[!t]
	\centering
	\includegraphics[width=13cm]{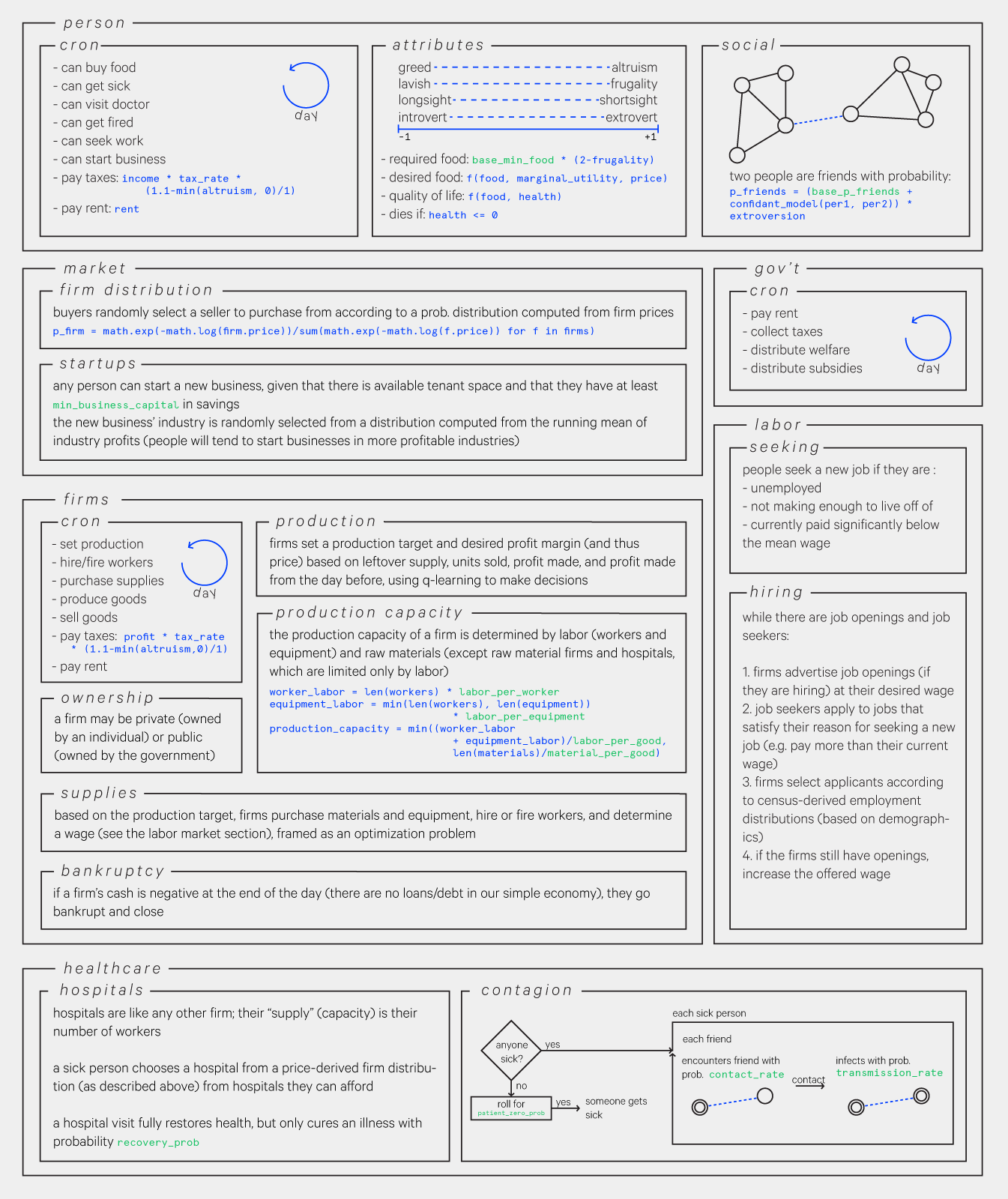}
	\caption{Processes of the model: the figures summarizes the process of the models. It describes the four-level attributes of the agents, their everyday activities and their social networking. It also contain main ideas for markets, their consumption process and startups generation, as well as firms’ details, production decisions and ownership. Finally, labor market hiring and firing procedures along with contagion and hospital descriptions.}
	\label{fig1}
\end{figure}

Our model integrates several components: individual persons, firms, and the government. In this model these entities typically interact via various markets. Each step of the simulation involves processing these various markets and adjusting each entity accordingly. A summary of the model is depicted in Figure \ref{fig1}.

\subsubsection{Person: citizen}

\paragraph{Attributes.}
As described previously, a simulated resident's demographics are generated from a Bayes' Net. 
\paragraph{Steps.}
At the start of a step, individuals decide whether or not they need to find a job or if they want to start a business.

Individuals only start a business if:
\begin{itemize}
	\item they do not already own a business
	\item there is a building with available space and affordable rent
	\item they can hire at least one employee
\end{itemize}
If these conditions are met, they start a business. The particular type of firm they decide on (see below for more details about the firms) is dependent on sectors’ profitability; i.e. they are most likely to pick a sector that has firms that are most profitable on average. Otherwise, if they do not already have a job, they seek out one. The dynamics of the labor market are discussed later.

Individuals then decide how much food they will purchase. First, they decide on the desired amount of food they want to purchase. This amount of food is computed based on a few factors: the minimum food necessary for survival, the marginal utility of excess food (considering price), and the amount of food currently stockpiled. Then they determine how much they can afford, and buy the lower amount of the two. 

If a person has less food than the survival amount at the end of a step, their health suffers. If they are unable to purchase food for a prolonged period of time, their health eventually goes below zero and they die.
If an individual is ill they may also decide to visit a hospital. The simulation’s healthcare system is described in further detail later.

\paragraph{Quality of life.}
To compute an individual’s quality of life we combine their perceived utility of the food they have and their current health, i.e.:

\begin{equation}
food*food\_utility+health\_utility(health)
\end{equation}

Quality of life is used as the primary signal of a person's well-being.

\subsubsection{Firms}

Our model consists of four types of firms:

\begin{enumerate}
	\item Hospitals
	\item Capital equipment firms
	\item Consumer good (food) firms
	\item Raw material firms
\end{enumerate}

Hospitals do not directly interact with any of the other firms. They treat sick people; the amount of sick people they are able to treat in a day is governed by the size of their staff.

Raw material firms produce materials which are necessary for both consumer good and capital equipment firms to produce their products. Capital equipment firms produce equipment which other firms (excluding hospitals) can purchase to increase the labor output of their workers. 

Finally, consumer good firms produce food which agents require to survive and may purchase in excess as well (as a luxury).

\paragraph{Production.}

The output of a firm is limited primarily by their available labor power, computed as a combination of employees and equipment. Firms may purchase equipment from capital equipment firms to increase labor power without hiring more workers. Each piece of equipment requires one worker to operate it.

With the exception of hospitals, each good requires some configurable labor cost to produce. Thus the production capacity of a firm - that is, how much a good they can produce in a simulation step - is limited by:

\begin{equation}
production\_capacity=labor\_power/labor\_cost\_per\_good
\end{equation}

Consumer good and capital equipment firms' production capacity is further limited by the amount of raw material they have on hand, i.e.:

\begin{dmath}
production\_capacity=min(labor\_power/labor\_cost\_per\_good, materials/material\_cost\_per\_good)
\end{dmath}

As previously mentioned, labor power is determined by both a firm’s workers and their available equipment. A piece of equipment increases labor power, but requires one worker to operate it. This system allows the model to explore the dynamics of automation.

For example, assume a good requires $10$ "labor" to produce, a single worker generates $20$ labor, and a piece of equipment adds $10$ labor. Say your firm wants to produce $10$ goods (requiring $100$ labor in total). With no equipment, five workers are required $5*20=100$. With four pieces of equipment, only four workers are required $4*(20+10) = 120$. In a technologically-advanced society, a piece of equipment may generate $50$ labor, in which case it would take only two workers and two pieces of equipment to produce an equivalent amount of goods.

\paragraph{Pricing \& Supply.}

Firms must decide what to price their goods and services at. Firms use Q Learning to make this decision, learning how to adjust their desired profit margin according to past sales performance in the market. In essence, based on past experience firms learn expected rewards from a profit margin adjustment, given other conditions in the economy. The main signals firms consider in this decision are how many products they sold the prior step, how much stock is leftover, and the change in profit.

Similarly, firms also use $Q$ Learning to set a desired production target for a step, determined from the same signals used to set the profit margin. Once a target is set, a firm then determines the optimum combination of human labor and equipment necessary to meet that target, purchasing or hiring employees as needed and as available.

\paragraph{Steps.}

Firms start a simulation step by setting a production target. They first consider their profit from the last step and any leftover supply. The $Q$ Learning algorithm returns a desired supply for this step, as well as the profit margin to set for this step.

Firms then evaluate employee salaries against the mean wage; in order to reduce production costs they fire employees that are being paid excessively above this mean. Similarly, they determine how much labor is required to produce the desired supply and compute the equipment and change in workers necessary to optimally (in terms of cost effectiveness) achieve this labor output. If the desired change in workers is negative, they randomly fire workers based on demographics and an employment distribution. If the desired change in workers is positive, firms also compute a desired wage and release “job openings” into the labor market.

Firms which require raw materials also decide on how much of that to purchase to produce their desired supply.

After interacting with the necessary markets (e.g. buying equipment from the capital equipment market, buying materials from the raw materials market; see below for more details), firms re-assess what they were able to acquire (in terms of new hires, capital equipment, and raw materials) and produce as much as they can (up to their desired production target).

Firms then set their price (based on cost-per-unit and desired profit margin) and go to their respective market.

\subsubsection{Government}

\paragraph{Steps.}

Like firms, the government also uses a $Q$ Learning algorithm, but adjusts tax rate and welfare payments instead of production targets and profit margins. The signal the government uses to determine whether or not such adjustments are "successful" is the mean quality of life of its citizens.

Each step the government also collects taxes from firms and individuals based on the current tax rate. Depending on player voting, the government may also distribute subsidies to the different industries.

\paragraph{Voting.}

Participants may join the simulation remotely via a web browser. Upon joining the simulation, they are assigned a random citizen. Players may periodically propose and/or vote on legislation in order to influence the government's decisions and in this way influence the simulation outcomes as well.

In the present version, a random citizen is chosen to propose legislation. Through these legislations, players may nationalize or privatize industries, adjust welfare payments, welfare threshold, tax rate, or industry subsidies. By voting players can attempt to shape the world in a way that benefits their assigned citizen.

\subsubsection{Markets}

\paragraph{Firms.}

Production firms (i.e. capital equipment, raw material, and consumer good firms) buy and sell from one another in markets. Purchasing firms select which supplying firm to buy from by random choice, such that supplying firms with the lowest prices are most likely to be selected. Once a firm runs out of supply, they leave the market.

\paragraph{Labor. Seeking.} 

Individuals seek a job under the following conditions:
\begin{itemize}
	\item if, according to the mean consumer good price, they cannot afford the food they need to live on their current wage (which is $0$ if they are unemployed).
	\item if their current wage is below the mean wage by some configured amount; that is, they believe to be underpaid.
\end{itemize}

Individuals never seek a job if they are a business owner.

\paragraph{Labor. Hiring.}

Firms announce job openings as described above. Individuals only apply to a job if it at least pays enough for them to purchase the bare minimum food for survival. Firms then consider applicants, hiring employees randomly. This also takes into account job referrals, where an applicant has a greater chance of being hired if they know someone presently employed at the firm.

\subsubsection{Healthcare}

Our model includes a simple healthcare system. Hospitals are represented as another firm in the economy and pricing is determined in a way similar to other firms (i.e. using $Q$ Learning to set a profit margin).

Individuals may get sick, in which case their health starts decreasing. An individual decides whether or not they visit a hospital based on their frugality, the perceived change in utility of getting treated (that is, the sicker they are, the more likely they are to visit a hospital), and the affordability of treatment. Treatment does not guarantee that the individual is healed of their illness; they may need to receive additional treatment.

Illness may propagate throughout the world via social networks using a simple contagion model. That is, a friend of someone ill may themselves get sick depending on contact rate and transmission rate parameters.

\subsubsection{Parameters}

It is inevitable that any simulation makes many assumptions about how the world works. We have tried to make many of these assumptions modifiable as accessible configuration parameters.

The parameters are: consumer\_good\_utility, n\_buildings,  labor\_cost\_per\_good, base\_min\_consumption, 
labor\_per\_equipment, labor\_per\_worker, transmission\_rate, 
wage\_under\_market\_multiplier, residence\_size\_limit,  tax\_rate\_increment, \\
profit\_increment, recovery\_prob,  supply\_increment, min\_business\_capital,
welfare\_increment, welfare, max\_tenants, contact\_rate,  starting\_welfare\_req, extravagant\_wage\_range, tax\_rate,  starting\_wage, patient\_zero\_prob, rent, wage\_increment, 
material\_cost\_per\_good, sickness\_severity.

These act as levers on various parts of the model. For example, "transmission\_rate", "contact\_rate", and "recovery\_prob" are standard parameters of a contagion model and affect how illness spreads throughout the world. Parameters such as "labor\_cost\_per\_good", "labor\_per\_worker", and "labor\_per\_equipment" affect how difficult it is to produce goods and greatly influence the dynamics of firm production and the economy. 

Parameters such as "wage\_under\_market\_multiplier" and "extravagant\_wage\_range" affect the dynamics of the labor market, determining when firms consider an employee overpaid and when employees consider themselves underpaid.

In theory, anyone can easily adjust these parameters to match their own assumptions. We also open them up in a few pre-configured settings ("scenarios") as detailed below to facilitate speculative exploration of the model.

\subsection{Scenarios}

Players – or rather, modelers - can choose from three levels of intensity on scenarios along the axes of food, technology, and disease, resulting on 27 possible combinations.
\begin{itemize}
	\item Food
	\begin{itemize}
		\item a bioengineered super-nutritional food is available
		\item "regular" food is available
		\item a blight leaves only poorly nutritional food
	\end{itemize}
	\item Technology
	\begin{itemize}
		\item hyper-productive equipment is available
		\item "regular" technology is available
		\item a massive solar flare from the sun disables all electronic equipment
	\end{itemize}
	\item Disease
	\begin{itemize}
		\item disease has been totally eliminated
		\item "regular" disease
		\item an extremely infectious and severe disease lurks
	\end{itemize}
\end{itemize}

\subsection{Inputs}

Data and parameters were retrieved from social and economic microdata, especially from IPUMS USA , New York Department of Labor and S\&P. In detail, we list the following original repositories: 

\begin{itemize}
	\item Frequently Occurring Surnames from the Census 2000. Surnames occurring $>= 100$ more times in the 2000 census \footnote{\url{http://www2.census.gov/topics/genealogy/2000surnames/surnames.pdf}}
	\item Female/male first names from the Census 1990 \footnote{\url{http://deron.meranda.us/data/}}
	\item Household and individual IPUMS data for 2005-2014 from IPUMS USA, Minnesota Population Center, University of Minnesota \footnote{\url{https://usa.ipums.org/usa/index.shtml}}
	\item PUMS network map of NY, hand-compiled from \footnote{\url{http://www.nyc.gov/html/dcp/pdf/census/puma\_cd\_map.pdf}}
	\item NYC unemployment data was retrieved from New York State Department of Labor \footnote{https://labor.ny.gov/stats/laus.asp}
	\item S\&P500 data was retrieved from Open Knowledge's Standard and Poor's (S\&P) 500 Index Data including Dividend, Earnings and P/E Ratio \footnote{\url{http://data.okfn.org/data/core/s-and-p-500}}
	\item Friendship model parameters were taken from \cite{smith}
	\item Annual expenses were taken from Living Wage Calculator \footnote{\url{http://livingwage.mit.medu/counties/36061}} made available by Amy K. Glasmeier, Carey Anne Nadeau and Eric Schultheis.
	
\end{itemize}

\section{Outputs and visualization}\label{outputs}

\subsection{Visualizing disparity}

\begin{figure}[!t]
	\centering
	\includegraphics[width=10cm]{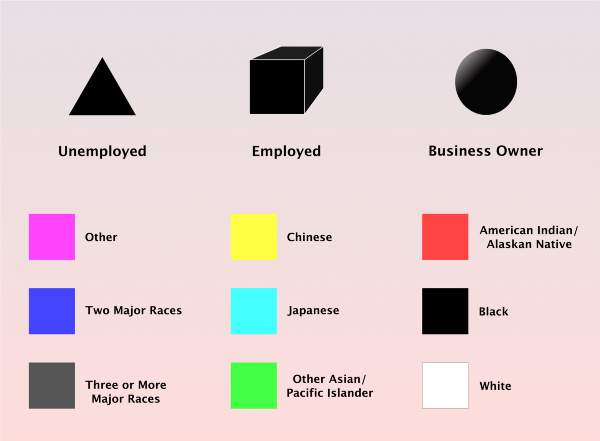}
	\caption{Representation of the citizens of the model depicting ethnicity and professional status.}
	\label{fig2}
\end{figure}

In order to visualize how different markets, sectors and ethnic population are changing in real time a form scheme is used. For the labor market pyramids are unemployed, cubes are employed, and spheres are business owners.  The simulant shapes are then colored according to Census race categories Figure \ref{fig2}. 

\begin{figure}[!t]
	\centering
	\includegraphics[width=10cm]{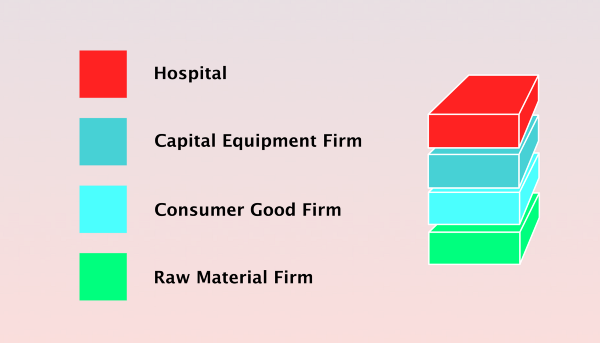}
	\caption{Representation of the buildings of the model according to industrial sectors.}
	\label{fig3}
\end{figure}

\begin{figure}[!t]
	\centering
	\includegraphics[width=10cm]{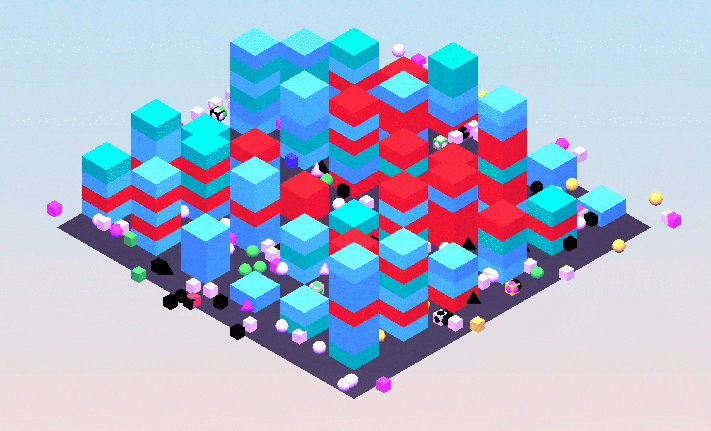}
	\caption{A snapshot of the visualization of the model with buildings and agents which are constantly moving in a dynamic fashion.}
	\label{fig4}
\end{figure}

The city's buildings also convey some information - each rectangular slice represents a different business, and each color corresponds to a different industry (raw material firms, consumer good firms, capital equipment firms, and hospitals) Figure \ref{fig3}. The shifting colors and height of the city becomes an indicator of economic health and priority - as sickness spreads, hospitals spring up accordingly Figure \ref{fig4}.

\subsection{Data output: illustration}

As an illustrative result, we present in Figure 5 some exploratory results for four scenarios. The first scenario, named 'Positive', refers to the top choices of the available alternatives, i.e., nutritious food, high productivity and no diseases. The second scenario, 'Low technology', maintains the beneficial food and the lack of diseases, but drops the level of productivity to no electronic equipment. An 'Average' scenario corresponds to regular quality food, technology and healthy. Finally, a negative scenario is made up of poor quality food, low technology and a highly infectious environment. 

All results are unbalanced with high levels of bankruptcies and losses. The ‘Negative’ scenario shows the worst quality of life with low increase in prices, wages and profits as the economy experiences very low trade and general activity. The 'Average' scenario, in turn, leads to very high prices of materials, a high peak in wages, both leading to high and varying consumer’s prices and low quality of life. The 'Positive' scenario also does not seem very favorable with high and early number of bankruptcies and constant increasing in wages. Further, a continuous downward trend in quality of life is observed, although better than the 'Negative' scenario.

The scenario with 'Low technology' seems to be the most reasonable one with an initial decrease in quality of life that rebounds in the later run of the model. That seems to happen, as prices remain much lower than those observed in the 'Average' and 'Positive' scenarios, although not at zero as the 'Negative' one. 

Overall, this preliminary exercise illustrates that further calibration and certainly validation are necessary before applying the model for public policy analysis. The aim of this text, however, is restricted to the presentation and discussion of the platform as a needed arena for city simulation debate. 

\begin{figure}
	\centering
	\subfloat[Quality of life]{{\includegraphics[width=5.7cm]{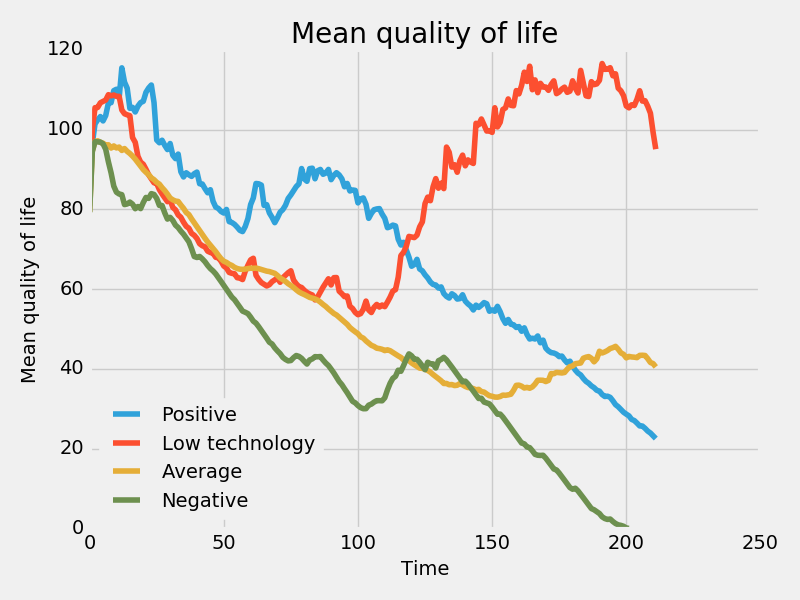} }}
	\quad
	\subfloat[Bankuptcies]{{\includegraphics[width=5.7cm]{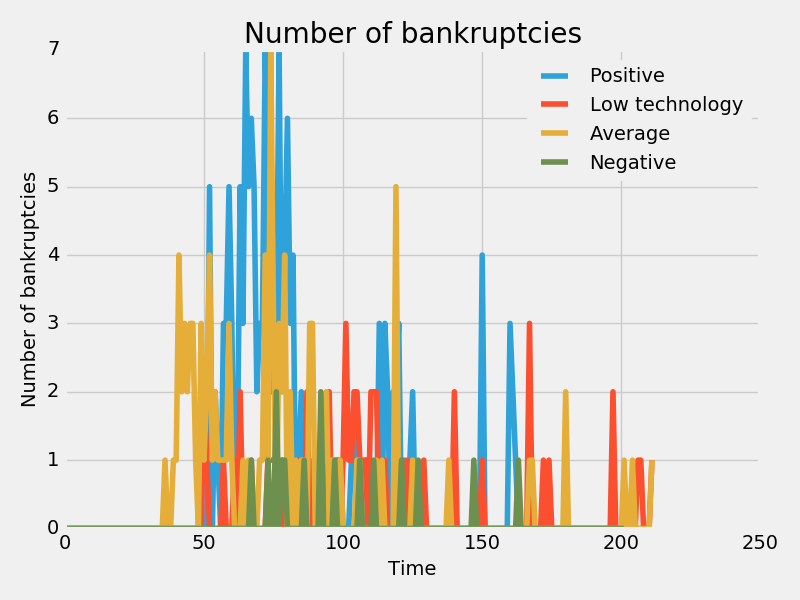} }}
	\quad
	\subfloat[Material prices]{{\includegraphics[width=5.7cm]{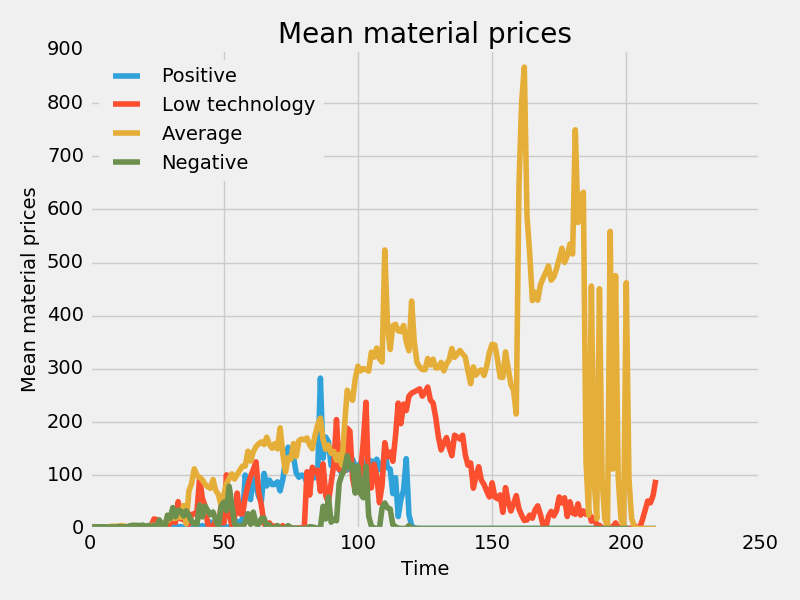} }}
	\quad
	\subfloat[Wage]{{\includegraphics[width=5.7cm]{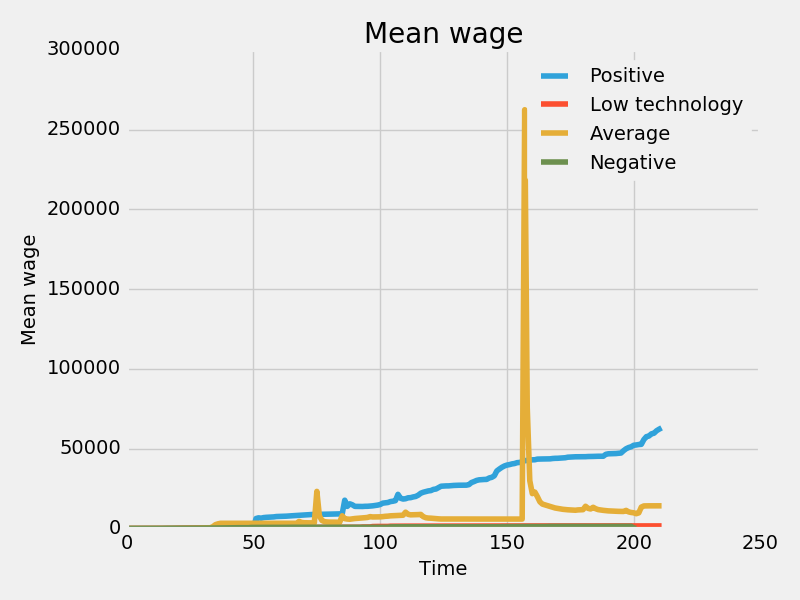} }}
	\quad
	\subfloat[Consumers goods' profit]{{\includegraphics[width=5.7cm]{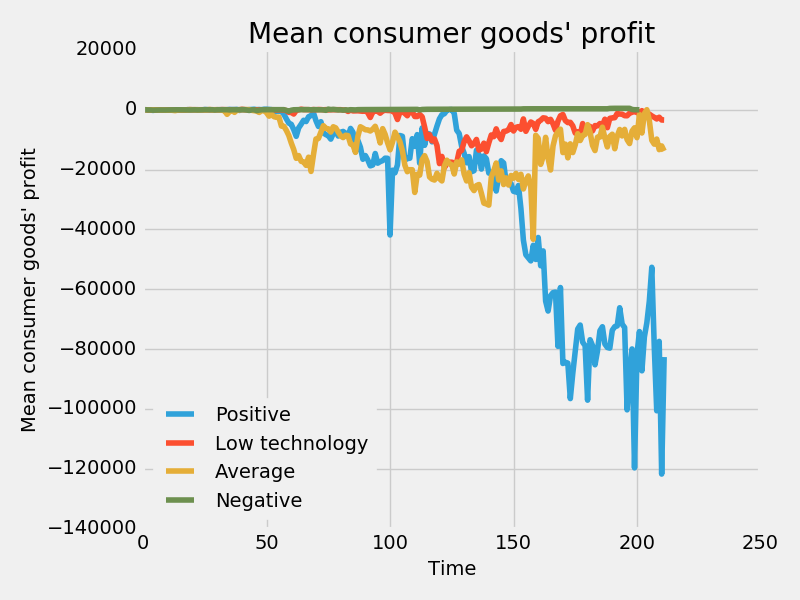} }}
	\quad
	\subfloat[Consumers goods' price]{{\includegraphics[width=5.7cm]{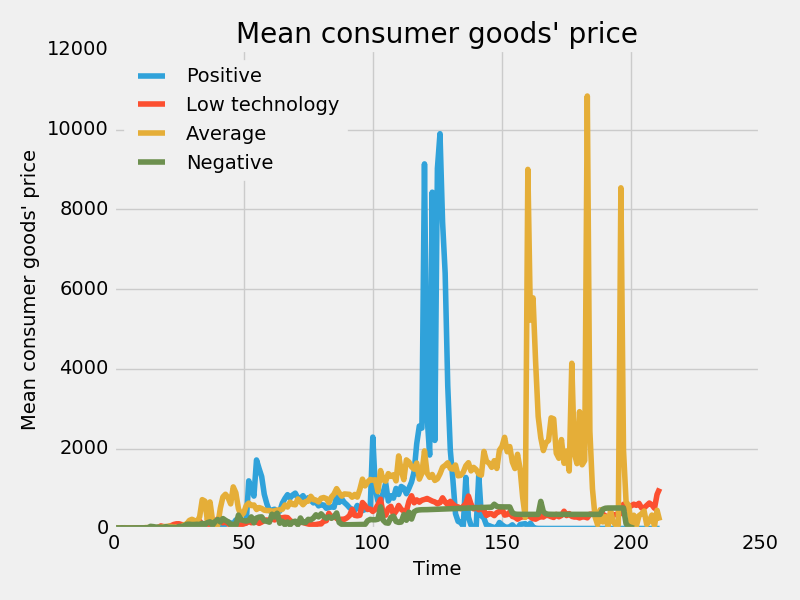} }}
	\caption{Results of four alternative scenarios with varying levels of food quality, productivity and health conditions. Most reasonable scenario seems to be 'Low technology' with high quality food and health and low productivity.}
	\label{fig5}
\end{figure}

\section{Final Considerations}\label{final}

City life is a palimpsest – an accumulated of experiences, dynamics and interactions. Forecasting or predicting results of policies on such an ever-changing masse of information is a daunting task. That does not mean that city interactions should not be studied and shed light upon. Better yet if such unraveling of city life is accompanied by rich visualization and interaction of modelers and prospect users. 

This paper presents a general model of city life simulation. It uses mixed methodologies, applied to an agent-based model to integrate citizens and cities daily activities within one simulation alone. Businesses, people and government decision processes are described. 

The simulation itself is portrayed as an exploratory initial framework to subside future developments. Its contribution lays on the amalgam of different elements united together in a single simulation. The present model is limited by its capacity to replicate the economics in the long run and further adjustments of parameters may be necessary. Yet, we believe the offered model is a sufficient novel proposal to foster city life analysis and understanding and thus bettering city life. 


\end{document}